\documentclass[aps,prd,twocolumn,preprintnumbers,nofootinbib,superscriptaddress,amsmath, longbibliography]{revtex4-2}
\usepackage{graphicx}
\usepackage{amssymb}
\usepackage{amsfonts}
\usepackage{url}
\usepackage[utf8]{inputenc}
\usepackage[dvipsnames]{xcolor}
\usepackage[breaklinks=true,colorlinks=true,
linkcolor=blue,urlcolor=Magenta,citecolor=Magenta,
bookmarks=true, bookmarksopenlevel=2]{hyperref}
\usepackage{bm}
\usepackage{array}
\usepackage{multirow}
\usepackage[normalem]{ulem}
\usepackage{longtable}
\usepackage{epsfig}
\usepackage{float}

\newcommand{\be}{\begin{equation}}
\newcommand{\ee}{\end{equation}}
\newcommand{\bea}{\begin{eqnarray}}
\newcommand{\eea}{\end{eqnarray}}
\newcommand{\nn}{\nonumber}

\newcommand{\Geff}{$G_{\rm eff}$ }

\newcommand{\ie}{{\it i.e.} }

\renewcommand{\[}{\begin{equation}}
\renewcommand{\]}{\end{equation}}

\def\lcdm{$\Lambda$CDM }

\usepackage{array}
\makeatother

\begin{document}

\preprint{IFT-UAM/CSIC-22-112}

\title{Machine learning constraints on deviations from general relativity from the large scale structure of the Universe}

\author{George Alestas}
\email{g.alestas@csic.es}
\affiliation{Instituto de F\'isica Te\'orica UAM-CSIC, Universidad Auton\'oma de Madrid, Cantoblanco, 28049 Madrid, Spain}
\affiliation{Department of Physics, University of Ioannina, GR-45110, Ioannina, Greece}

\author{Lavrentios Kazantzidis}
\email{l.c.kazantzidis@gmail.com}
\affiliation{Department of Physics, University of Ioannina, GR-45110, Ioannina, Greece}

\author{Savvas Nesseris}
\email{savvas.nesseris@csic.es}
\affiliation{Instituto de F\'isica Te\'orica UAM-CSIC, Universidad Auton\'oma de Madrid, Cantoblanco, 28049 Madrid, Spain}

\date{\today}

\begin{abstract}
We use a particular machine learning approach, called the genetic algorithms (GA), in order to place constraints on deviations from general relativity (GR) via a possible evolution of Newton's constant $\mu\equiv G_\mathrm{eff}/G_\mathrm{N}$ and of the dark energy anisotropic stress $\eta$, both defined to be equal to one in GR. Specifically, we use a plethora of background and linear-order perturbations data, such as type Ia supernovae, baryon acoustic oscillations, cosmic chronometers, redshift space distortions and $E_g$ data. We find that although the GA is affected by the lower quality of the currently available data, especially from the $E_g$ data, the reconstruction of Newton's constant is consistent with a constant value within the errors. On the other hand, the anisotropic stress deviates strongly from unity due to the sparsity and the systematics of the $E_g$ data. Finally, we also create synthetic data based on a next-generation survey and forecast the limits of any possible detection of deviations from GR. In particular, we use two fiducial models: one based on the cosmological constant $\Lambda$CDM model and another on a model with an evolving Newton's constant, dubbed $\mu$CDM. We find that the GA reconstructions of $\mu(z)$ and $\eta(z)$ can be constrained to within a few percent of the fiducial models and in the case of the $\mu$CDM mocks, they can also provide a strong detection of several $\sigma$s, thus demonstrating the utility of the GA reconstruction approach.
\end{abstract}

\maketitle

\section{Introduction \label{sec:intro}}
Currently the de facto cosmological model is widely considered to be the $\Lambda$ Cold Dark Matter ($\Lambda$CDM) scenario, which is based on general relativity (GR) and it contains not only ordinary matter but also incorporates two other ingredients that are the backbone of this model. The cosmological constant $\Lambda$, is crucial for the description of the accelerated expansion of the Universe, as well as the Cold Dark Matter (CDM). Ever since the observed accelerated expansion of the Universe it holds, almost unanimously, the scepters in the field of Cosmology. Lately, however, there is mounting evidence supporting the idea that perhaps \lcdm is not infallible. 

To that end, there have been observed a number of discrepancies/tensions (for comprehensive reviews see Refs. \cite{Perivolaropoulos:2021jda, CANTATA:2021ktz, Abdalla:2022yfr}) that emerge when comparing the predictions made by the model to observational data, such as the $S_8$ tension \cite{Joudaki:2017zdt, DES:2017myr, Basilakos:2017rgc, Nesseris:2017vor, Kazantzidis:2018rnb, Perivolaropoulos:2019vkb, Skara:2019usd, Alestas:2021xes}, cosmic microwave background (CMB) anisotropy anomalies \cite{Rassat:2014yna, Schwarz:2015cma, Planck:2019evm, Perivolaropoulos:2014lua, Muir:2018hjv, Bayer:2020pva}, the lithium problem \cite{Cyburt:2003fe, Cyburt:2008kw, Poulin:2015woa, Mori:2019cfo, Hayakawa:2020bjr, Ishikawa:2020fbm}, etc. Arguably though, the most well-known of these tensions is the so-called Hubble tension \cite{Ishak:2018his, Kazantzidis:2019nuh, DiValentino:2021izs, Perivolaropoulos:2021jda, CANTATA:2021ktz, Schoneberg:2021qvd, Alestas:2020mvb, Shah:2021onj, Abdalla:2022yfr}. This tension involves the $5\sigma$ disagreement between the value of $H_0$ as given by the Planck collaboration via the CMB data ($H_0=67.27 \pm 0.60 \; \mathrm{km}\; \mathrm{s}^{-1}\; \mathrm{Mpc}^{-1}$) \cite{Planck:2018vyg}, and the one given by the SH0ES team using SnIa data calibrated by Cepheids ($H_0=73.04 \pm 1.04 \; \mathrm{km}\; \mathrm{s}^{-1}\; \mathrm{Mpc}^{-1}$) \cite{Riess:2021jrx}. 

If these tensions are not due to some unknown systematic effect, then some of them might suggest that \lcdm needs to be further modified. As a result, a number of possible early or/and late time solutions to these tensions have been proposed in the literature, including a variety of modified gravity models. A number of recent works \cite{Alestas:2020zol, Marra:2021fvf, Alestas:2021nmi, Perivolaropoulos:2021bds, Alestas:2022xxm, Perivolaropoulos:2022txg, Odintsov:2022eqm} in the field argue that in order to reconcile the Hubble and growth tensions, an evolution of Newton's constant $G_{\rm eff}$ which leads to its decrease as the Universe evolves is needed. More specifically, this modified gravity solution involves ultra-late time gravitational transitions that in their core, explore and enable the possibility of a transition in the SnIa absolute magnitude $M_B$ which would allow for the easing of both the Hubble and $S_8$ tensions, whilst leaving the standard cosmology intact within the ultra-low redshift range $z \in [0.01, 1000]$. 

This transition in $M_B$ could very well be a product of a gravitational transition of $G_{\rm eff}$, since the two are connected via the Chandrasekhar mass $m_{ch}$ that evolves as $m_\mathrm{ch} \sim G_{\rm eff}^{-3/2}$, leading to $\Delta M_B = \frac{15}{4}\log \mu$ \cite{Amendola:1999vu, Gaztanaga:2001fh}, where 
\be
\mu \equiv \frac{G_{\rm eff}}{G_{\rm N}},\label{eq:mu}
\ee
and $G_{\rm N}$ is the bare Newton's constant as measured by Cavendish-type experiments in a laboratory.

Within the context of this paper, we use Machine Learning (ML) in the form of Genetic Algorithms (GA) in order to assess whether current cosmological data from large scale structure (LSS) probes predict an evolution of the $G_{\rm eff}$. The advantage of ML methods, in this case, is that they allow for a bottom-up reconstruction of $G_{\rm eff}$ solely from the data, avoiding theoretical biases and enabling us to look for features that may otherwise not be readily detected via more traditional approaches \cite{Arjona:2020kco}.

More specifically, we use the Pantheon sample of SnIa data, an up-to-date collection of Baryon Acoustic Oscillations (BAO) data, Cosmic Chronometer (CC) data, an up-to-date compilation of growth rate/redshift space distortions data, and a similarly up-to-date collection of $E_g$ data. Via the GA process, we are able to ascertain the best-fit functions for the luminosity distance $D_L(z)$ using the Pantheon SnIa data, the angular diameter distance $D_A(z)$ using the BAO data, the Hubble parameter $H(z)$ using the CC data, the growth rate of perturbations $f\sigma_8(z)$ via the growth rate data and the $P_2(z)$ observable, which is a model-independent probe of modified gravity defined as the ratio of the lensing function $\Sigma$ multiplied by $\Omega_\mathrm{m,0}$ and the growth rate $f$ \cite{Zhang:2007nk, Amendola:2012ky}, using the $E_g$ data. The combination of these cosmological quantities allows us to probe the possible evolution of Eq.~\eqref{eq:mu}.

In linear order perturbation theory one more quantity can also be defined in order to parameterize deviations from GR, namely the anisotropic stress $\eta$. The latter is equal to the ratio of the Newtonian potentials and, in the absence of any matter-induced anisotropic stresses, is equal to unity. Here, we also use the GA in order to reconstruct $\eta$ and we compare those results with an analysis using earlier data by Ref.~\cite{Arjona:2020kco}.

Having done the GA reconstructions with the currently available data, we also forecast the constraining power of approach using mock data based on a future next-generation survey like Euclid or DESI, albeit without targeting any of them specifically, in order to determine how accurately forthcoming surveys will be able to constrain deviations from GR. To do so, we use two mocks, one based on \lcdm and one more, using parametrizations for $\mu(z)$ and $\eta(z)$.

The structure of our paper is as follows: in Sec.~\ref{sec:ga} we present a brief summary of the GA approach and how it works, in Sec.~\ref{sec:data} we present the compilations of currently available and mock data we use, while in Sec.~\ref{sec:results} we present the results of our analysis in the cases of both the currently available and mock data. Finally, in Sec.~\ref{sec:conclusions} we present our conclusions.

\section{Theory \label{sec:theory}}
Here we describe in more detail the theoretical setup of our analysis, which is based on cosmological perturbation theory, setting the stage for the machine learning reconstructions in later sections. We mainly follow Refs.~\cite{Pinho:2018unz, Skara:2019usd, Arjona:2020kco}, albeit with slight changes in the notation.

First, we assume a perturbed flat Friedmann-Lemaitre-Robertson-Walker (FLRW) metric
\be
ds^2=-(1+2\Psi)\mathrm{d} t^2 + a(t)^2(1-2\Phi)\mathrm{d} \vec{x}{}^2,\label{eq:FLRW}
\ee
where $a(t)$ is the scale factor given in terms of the cosmic time $t$, while $\Phi$ and $\Psi$ are the scalar potentials in the Newtonian gauge. Using the field equations it is possible to show that for a plethora of different modified gravity models, in the subhorizon and quasistatic regime, the Newtonian potentials obey the following Poisson equations in Fourier space \cite{Tsujikawa:2007gd}
\bea
-\frac{k^2}{a^2}\Psi &= 4\pi G_{\rm N} \mu(k,a)\rho_m\delta_m,\label{eq:pois1}\\
-\frac{k^2}{a^2}(\Psi+\Phi) &= 4\pi G_{\rm N} \Sigma(k,a)\rho_m\delta_m,\label{eq:pois2}
\eea
where $k$ is the wavenumber, $\rho_m$ is the matter density, $\delta_m\equiv \frac{\delta \rho_m}{\rho_m}$ is the growth factor of matter density perturbations and the parameters $\Sigma$ and $\mu$ characterize deviations from GR. In particular, in GR the corresponding limits are $\Sigma=2$ and $\mu=1$, see Refs.~\cite{Ma:1995ey,Pinho:2018unz, Arjona:2020kco}, while in other theories they can be in general scale and time-dependent \cite{Tsujikawa:2007gd}.

Furthermore, modified gravity models in general  may also induce an anisotropic stress, which can be parameterized by the parameter $\eta$ defined as the ratio of the two potentials
\be 
\eta\equiv \frac{\Phi}{\Psi}, \label{eq:aniso}
\ee 
which in GR has the limit $\eta=1$ as expected in the absence of anisotropic stresses from matter (e.g. from neutrinos or photons \cite{Ma:1995ey}). The anisotropic stress or gravitational slip quantity $\eta(z)$ is both an element that allows for the non-minimal coupling of the dark energy with gravity in the context of the Jordan frame, and a purely geometric characteristic of higher order modified gravity models \cite{Saltas:2011xlz, Amendola:2013qna, Sawicki:2012re}. 

The evolution of the cosmological overdensity $\delta_m(a)$ satisfies, in the absence of neutrinos and for most modified gravity theories, the following differential equation in the subhorizon regime
\be 
\delta_m''(a)+\left(\frac{3}{a}+\frac{H'(a)}{H(a)}\right)\delta_m'(a)-\frac{3}{2}\frac{\Omega_{\textrm{m,0}} \, \mu(k,a)}{a^5 H^2(a)/H_0^2} \,\delta_m(a)=0, \label{eq:odedeltaa}
\ee
where primes denote differentiation with respect to $a$ and $k$ corresponds to the scale. Observations strongly suggest the existence of a large-scale structure of the Universe that was created during a matter domination era, typically assumed to be $z\in[1,10^4]$. It is then easy to show that the growth should behave as $\delta(a\ll 1)\sim a$ at an initial time deep in matter domination.

A related quantity is the growth rate $f$ which is defined as 
\be 
f(a) \equiv \frac{\mathrm{d} \ln \delta_m (a)}{\mathrm{d} \ln a} \label{eq:fadef},
\ee
and is a proxy for the growth of matter density perturbations on large scales. However, in the past two decades the vast majority of LSS surveys report instead the bias-independent product $f\sigma_8(a)=f(a) \cdot \sigma_8(a)$, where 
\be 
\sigma_8(a) \equiv \frac{\sigma_8}{\delta_m(1)} \, \delta_m(a),\label{eq:sig}
\ee
with $\sigma_8$ corresponding to the density rms fluctuations within spheres of radius on scales of about $8 h^{-1} \, \rm{Mpc}$.

Therefore, given a Hubble rate $H(a)$ and a parametrization for $\mu(k,a)$, Eq.~\eqref{eq:odedeltaa} can be solved either numerically or analytically and using the solution, the theoretical prediction for $f\sigma_8$ is readily constructed via Eqs.~\eqref{eq:fadef}-\eqref{eq:sig}.

In order to reconstruct the aforementioned parameters $\mu$, $\Sigma$ and $\eta$, we follow Ref.~\cite{Pinho:2018unz} and we define a set of variables 
\bea
P_2(a)&=&\frac{\Sigma \, \Omega_{\textrm{m,0}}}{f}, \label{eq:P2}\\
P_3(a)&=&\frac{\mathrm{d} \ln f\sigma_8}{\mathrm{d} \ln a}, \\
E(a)&=& H(a)/H_0,
\eea
where $\Omega_{\textrm{m,0}}$ is the present value of the matter density parameter. Then, it follows that the anisotropic stress is given by the relation \cite{Pinho:2018unz}
\be
1+\eta(a)=\frac{3\,P_2(a)\,a^{-3}}{2 \, E(a)^2\left[P_3(a)+2+\frac{\mathrm{d} \ln E}{\mathrm{d} \ln a}\right]}. \label{eq:eta}
\ee
Combining Eqs.~\eqref{eq:pois1}-\eqref{eq:pois2} and \eqref{eq:aniso} we can find the expression   
\be 
\Sigma=\mu\,(1+\eta),
\ee 
which then from the definition of $P_2$ via Eq.~\eqref{eq:P2}, gives 
\be
\mu = \frac{f \, P_2}{(1+\eta) \, \Omega_{\textrm{m,0}}}.\label{eq:mufrac}
\ee
In the latter equation the final ingredient is the parameter $P_2$, which can be shown to be related to the so-called $E_g$ statistic, see Ref.~\cite{Pinho:2018unz} and references therein. In fact, $E_g$ is the expectation value of the ratio of lensing and galaxy clustering observables at a scale $k$
\be 
E_g = \Big\langle \frac{a\,\nabla^2 (\Psi+\Phi)}{3\,H_0\,f\,\delta_m}\Big\rangle_k,\label{eq:egstat}
\ee 
which after using the Poisson equation \eqref{eq:pois1}-\eqref{eq:pois2} reduces to $E_g=2P_2$. As $\Omega_{\textrm{m,0}}$ is not directly observable but can only inferred by other probes, we can re-write the Eqs. ~\eqref{eq:eta} and \eqref{eq:mufrac} in terms of the actual observables as
\begin{align}
    \Omega_{\textrm{m,0}}\,\mu(a) &= \frac{2\,f(a) \, E_g(a)}{1+\eta(a)},\label{eq:om_mu}\\
    1+\eta(a) &= \frac{3\,E_g(a)\,a^{-3}}{\, E(a)^2\left[P_3(a)+2+\frac{\mathrm{d} \ln E}{\mathrm{d} \ln a}\right]}, \label{eq:eta_obs}
\end{align} 
where in what follows we will consider the combination $\Omega_{\textrm{m,0}}\,\mu(a)$ as it is independent from $\Omega_{\textrm{m,0}}$. Still, any deviation of Eq.~\eqref{eq:om_mu} at any redshift from a constant value would be a smoking-gun signature for deviations from GR at late times \cite{Nesseris:2011pc}, so this makes it an extremely useful statistic.

\section{The Genetic Algorithms \label{sec:ga}}
In this section we describe the GA, as used in our work. The GA has had several applications in cosmology, see for example  Refs.~\cite{Bogdanos:2009ib,Nesseris:2010ep, Nesseris:2012tt,Nesseris:2013bia,Sapone:2014nna,Arjona:2020doi,Arjona:2020kco,Arjona:2019fwb}, in forecasts of future LSS surveys \cite{EUCLID:2020syl, Euclid:2021cfn, Euclid:2021frk}, but also in a wide range of areas such as particle physics \cite{Abel:2018ekz,Allanach:2004my,Akrami:2009hp}, astronomy and astrophysics \cite{wahde2001determination,Rajpaul:2012wu,Ho:2019zap} and other fields like computational science, economics, medicine and engineering \cite{affenzeller2009genetic,sivanandam2008genetic}. There are also several other similar symbolic regression methods, see for example \cite{Udrescu:2019mnk,Setyawati:2019xzw,vaddireddy2019feature,Liao:2019qoc,Belgacem:2019zzu,Li:2019kdj,Bernardini:2019bmd,Gomez-Valent:2019lny}.

The GA that we use in the present work in order to identify a possible evolution of \Geff correspond to an unsupervised symbolic regression of data process that in reality simulates the biological natural selection. Initially, a set of randomly selected orthogonal basis functions (which we denote as ``grammar") are chosen and are then subjected to crossover and mutation operations over time, until a specific termination criterion is met. For a set of vectors (or functions), orthogonality is a much stronger statement as it implies not only they are linearly independent, but also their dot product is zero. This is necessary in order to avoid degeneracies between the functions that could otherwise affect the convergence rate of the algorithm.

In particular, the initial ``grammar" population corresponds to the first generation of functions, in which we need to impose any appropriate conditions in order to obtain a physically meaningful result and applying the maximum likelihood method, a $\chi^2$ value is calculated for each member. Then,  convergence is achieved when  the $\chi^2$ does not improve or change for a few hundred generations. As can be seen in Fig.~1 of Ref.~\cite{Arjona:2019fwb}, when this happens it typically implies that the algorithm is at the minimum and it stays there, unlike an MCMC where the GA random-walker of a particular chain “oscillates” around it.

\begin{figure}[!t]
\centering
\includegraphics[width = 0.5\textwidth]{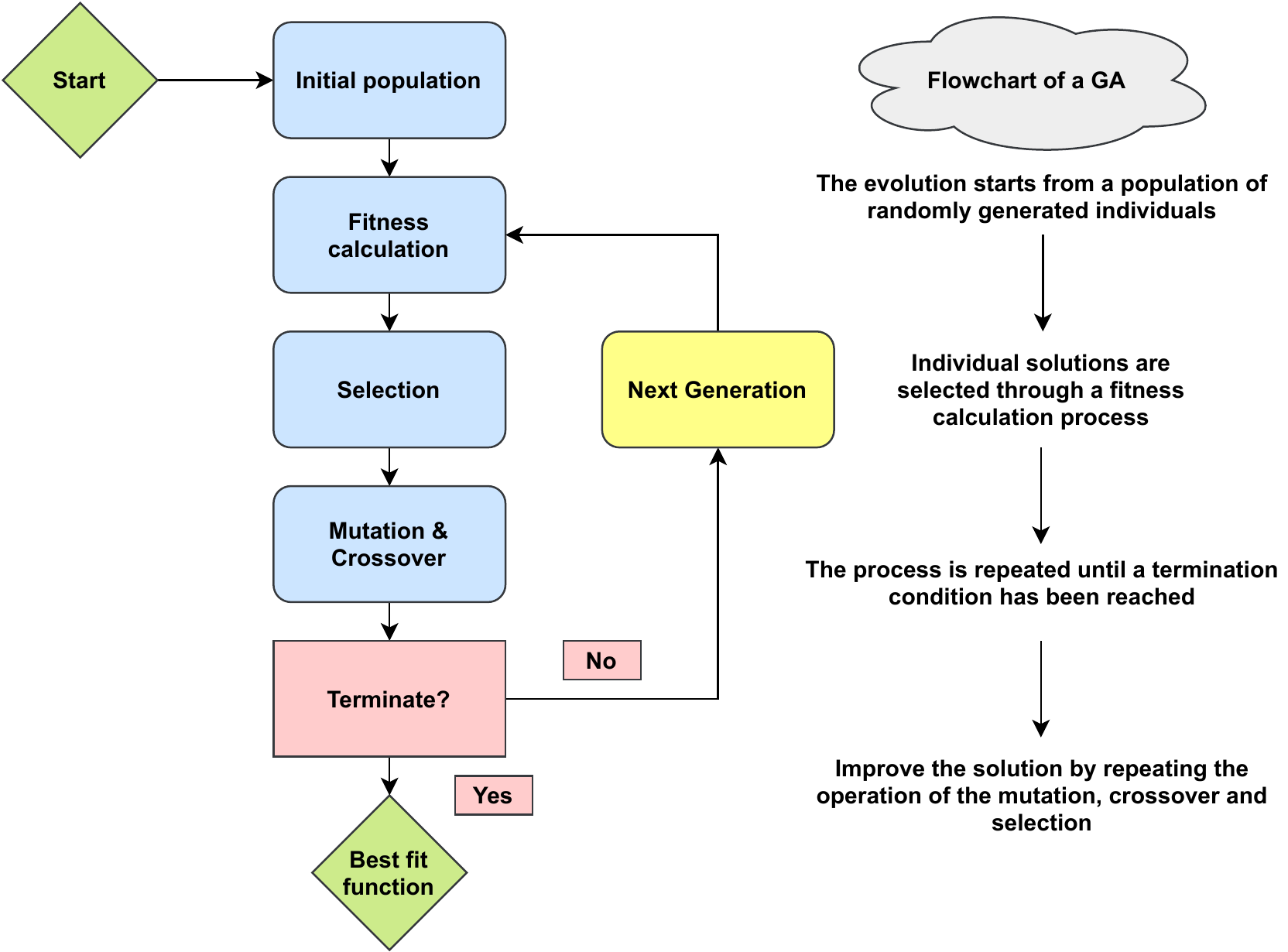}
\caption{Flowchart of a typical run of a Genetic Algorithm (from Ref.~\cite{Arjona:2020axn}). \label{fig:GA_flowchart}}
\end{figure}

Next, the best-fit functions from the first generation are determined through a tournament selection method, \ie a method that randomly chooses a subgroup of candidates and picks the dominant one from each subgroup. After that, the second generation of candidate functions is obtained by sequentially applying the crossover and mutation operations. Then this process is repeated a number of thousand times in order to guarantee convergence. Notice that with the term crossover we refer to the combination of two parties in order to form a descendant, while with the term mutation we describe the arbitrary change of an individual. A flowchart of a typical run of a GA can be seen in Fig.~\ref{fig:GA_flowchart}.

Naively one could think that grammar selection is crucial in order to obtain any physically meaningful results, however as it has been discussed in Ref.~\cite{Bogdanos:2009ib} the initial population of the grammar is irrelevant to the outcome and only influences the convergence rate. Regarding the termination criterion as it has been extensively discussed in Ref.~\cite{Bogdanos:2009ib} a number of different conditions can be applied. In this paper we let the GA reach the maximum number of generations and choose the best candidate based on a normal $\chi^2$ statistic.

As we have already mentioned one of the advantages of such a method is its model-independent non-parametric nature. As a result, the final outcome of the process corresponds to a group of continuous functions with respect to a variable $x$ (in our case we choose the variable to be the redshift $z$) and not to specific best-fit parameters of a model. However, this advantage can also be considered as a simultaneous drawback of the method since the lack of parameters leads to the abandonment of standard techniques for the calculation/propagation of the errors. Fortunately, this shortcoming can be bypassed through the path integral approach, as discussed in Ref.~\cite{Nesseris:2012tt}.

In a nutshell, as the GA explores the whole functional space, it effectively (in principle) passes through every possible value of the reconstructed value $f(x)$ at each $x$. This is effectively the same as doing a path integral and normalizing the likelihood reduces to a set of Gaussian integrals which result to a sum of error functions, see Ref.~\cite{Nesseris:2012tt}. This approach at determining the errors from the GA has been well-tested against bootstrap Monte Carlo and fisher matrix techniques and has been found to be in good agreement. Furthermore, it has the added advantage that it is much faster compared to bootstrap and requires less computational power.

\section{The data \label{sec:data}}
As we discussed in Sec.~\ref{sec:intro}, in order to constrain a possible evolution of the evolving Newton's constant \Geff through the GA, we use up-to-date cosmological data compilations, including SnIa, BAO and CC points, as well as growth and $E_g$ data. Note, we do not use any CMB data (even though it would have decreased the errors) as that would leave a huge gap in the points between $z\simeq2$ and $z\simeq1090$ that we would have to interpolate in-between. The exact compilations and the details for the likelihoods are described in what follows. 

\subsection{Type Ia Supernovae - The Pantheon Sample}
The SnIa dataset that we use in this work is the Pantheon sample \cite{Pan-STARRS1:2017jku}. This sample corresponds to one of the largest compilations of SnIa, which includes data both in the high and low redshift regimes that are observed by six different surveys. As a result, the sample incorporates 1048 SnIa in the redshift range $z \in (0.01,2.26]$\footnote{The newer compilation of the Pantheon+ was made publicly available in the final stages of this work, but we have not included it as we do not expect it to significantly affect our analysis.}. In general, SnIa are used extensively in the literature in order to measure the expansion rate $H(z)$ since they can be directly connected with the theoretically predicted apparent magnitude of a model $m_{th}(z)$ through the following equation 
\be
m_{th}(z)=M_B+5 \, \log_{10} \left[\frac{D_L(z)}{\rm{Mpc}} \right]+25 \label{eq:mthz},
\ee
with $M_B$ corresponding to the corrected absolute magnitude and $D_L$ is the luminosity distance, which in a flat FLRW Universe is given by
\be 
D_L(z)=\frac{c}{H_0}\,(1+z)\,\int_0^z\frac{1}{E(z')}\,\mathrm{d}z'.\label{eq:dLt}
\ee 

Even though some recent works \cite{Zhao:2019azy,Kazantzidis:2020tko,Sapone:2020wwz,Kazantzidis:2020xta,Dainotti:2021pqg} argue that the degenerate combination of this quantity with $H_0$ that appears when Eq. \eqref{eq:mthz} is expressed using the Hubble free luminosity distance $D_L(z)$ may contain useful information, normally the absolute magnitude $M_B$ is treated as a nuisance parameter and is marginalized \cite{SNLS:2011lii,SDSS:2014iwm,Pan-STARRS1:2017jku} following the method discussed in the Appendix C of Ref. \cite{SDSS:2014iwm}. So, in the present work we adopt the standard method and marginalize over $M_B$, leading as a result to a $\chi^2$ function of the form \cite{Arjona:2020kco}
\be 
\chi^2_{\rm{SnIa}}=A-\frac{B^2}{E}+\ln \left(\frac{E}{2\pi} \right) \label{eq:chi2SnIa},
\ee
where 
\begin{align*}
    A &=\Delta \vec{m} \cdot \vec{C}^{-1} \cdot \Delta \vec{m}, \\
    B &=\Delta \vec{m} \cdot \vec{C}^{-1} \cdot \Delta \vec{I}, \\ 
    E &=\vec{I} \cdot \vec{C}^{-1} \cdot \vec{I}.
\end{align*}
In the above expression $\vec{C}$ is the total covariance matrix (including both the statistical and systematic uncertainties), $\vec{I}=(1,1,\ldots,1)$ and $\Delta m=m_{\rm{obs}}(z_i)-m_{th}(z)$.

The total covariance matrix $\vec{C}$ arises as the sum of two independent matrices \cite{Pan-STARRS1:2017jku}. The first one is a diagonal matrix that includes the statistical uncertainties of the apparent magnitudes $m_{\rm{obs}}$ of each SnIa that has the following form
\be 
  \begin{pmatrix}
    \sigma^2_{m_{\rm{obs},1}} & 0 & \dots & 0 \\
    0 & \sigma^2_{m_{\rm{obs},2}} & \dots & 0 \\
    \vdots & \vdots & \ddots & \vdots \\
    0 & 0 & \dots & \sigma^2_{m_{\rm{obs},1048}}
  \end{pmatrix},
\ee
while the second one is a non-diagonal matrix related to systematic uncertainties. Therefore, for a specific form for the Hubble rate $H(z)$, the theoretically predicted apparent magnitude can be produced using Eq. \eqref{eq:mthz} and then the quality of the fit can be estimated applying the maximum likelihood method, \ie minimizing Eq. \eqref{eq:chi2SnIa}.

\subsection{Baryon Acoustic Oscillations}
Regarding the BAO compilation used in the present analysis we employ data from different observational missions such as the 6dFGS \cite{Beutler:2011hx}, WiggleZ \cite{Blake:2012pj,Escamilla-Rivera:2016qwv}, DES \cite{DES:2021esc} and Lya \cite{duMasdesBourboux:2020pck} surveys as well as other type of BAO data from the fourth generation of the SDSS mission (SDSS-IV), \ie the extended baryon oscillation spectroscopic survey (eBOSS) \cite{eBOSS:2020yzd}. However, since we incorporate data from different missions we cannot construct a unique $\chi^2$ formula as we did in the SnIa case, because the different missions do not constrain the same observational quantities. In reality, they report different dependent combinations that are related to one another. So, before we establish the corresponding $\chi^2$ formula, we need to recall some basic definitions.

It is known that the BAO correspond to a relic from the prerecombination era of the Universe that imposes a characteristic scale occurring either as a peak in the galaxy correlation function or as damped oscillations in the power spectrum. The radius of the sound horizon at last scattering $r_s$ is defined as \cite{Aubourg:2014yra}
\be 
 r_s (z_d) \equiv \int_{z_{d}}^\infty \frac{c_s(z)}{H(z)}\,\mathrm{d}z, \label{eq:soundhordef}
\ee
where the lower limit of the integration corresponds to the drag redshift $z_d$, \ie a redshift shortly after recombination that can be either calculated through a numerical package such as \texttt{CAMB}  or through the approximate formula discussed in Ref. \cite{Eisenstein:1997ik} [see its Eq. (4)]. The numerator of \eqref{eq:soundhordef} describes the sound speed of the baryonic-photon fluid given as \cite{WMAP:2008lyn}
\be
    c_s(z)=\frac{c}{\sqrt{3 \left(1+\frac{3\Omega_\mathrm{b,0}}{4\Omega_\mathrm{\gamma,0}} \frac{1}{1+z}\right)}}, \label{eq:csdef}
\ee
where $\Omega_\mathrm{b,0}$ and $\Omega_\mathrm{\gamma,0}$ are the present day baryon and photon densities respectively.

Taking into account the physical angular diameter distance $D_A(z)$ that in the context of a flat Universe is given as 
\be 
D_A(z)= \frac{c}{1+z} \int_0^z \frac{\mathrm{d}z'}{H(z')} \label{eq:daBAO},
\ee
two different observational quantities that are widely used in the literature can be obtained. The first one corresponds to the comoving angular diameter distance $D_M$ that is given through the following simple formula \cite{Bautista:2017zgn}
\be 
D_M(z)=(1+z) D_A(z), \label{eq:dmBAO}
\ee
while the second one is the combination 
\be 
D_V(z)=\left[\frac{c \, z \, D_M^2(z)}{H(z)} \right]^{1/3}. \label{eq:dvBAO}
\ee
Even though, Eqs. \eqref{eq:dmBAO} and \eqref{eq:dvBAO} are the main observational quantities that are published by the BAO surveys, some surveys may also report the $D_H$ observable that is defined as  
\be 
 D_H(z) \equiv \frac{c}{H(z)}.
\ee

Now returning to the specific data compilation of the current work, we use the results of the following surveys:
\begin{itemize}
    \item The 6dFGS and WiggleZ missions that report the quantity $d_z \equiv r_s(z_d)/D_V(z)$ giving the following values
    \begin{center}
    \begin{tabular}{ |c|c|c|c|c|c|c| } 
    \hline
    Missions & $z$ & $d_z$ & $\sigma_{d_z}$ & Ref. & Date\\
    \hline
    6dFGS & $0.106$ & $0.336$ & $0.015$ & \cite{Beutler:2011hx} & June 2011\\
    \hline
    \multirow{3}{4em}{WiggleZ} & $0.44$ & $0.073$ & $0.031$ & \multirow{3}{3.3em}{\cite{Blake:2012pj,Escamilla-Rivera:2016qwv}} & \multirow{3}{5em}{April 2012}\\
    & $0.6$ & $0.0726$ & $0.0164$ & &\\
    & $0.73$ & $0.0592$ & $0.0185$ & &\\
    \hline
    \end{tabular}
    \end{center}
    which lead to a $\chi^2$ formula of the form 
    \be 
    \chi^2_{\rm{6dFS,WigZ}}=V^i \, C_{ij}^{-1} \, V^j.
    \ee
    In this case, the vector $V^i$ is given as $V^i=d_{z,i}-d_z(z_i)$, whereas the covariance  matrix $C_{ij}$ yields 
    \be 
        \begin{pmatrix}
        \frac{1}{0.015^2} & 0 & 0 & 0 \\
        0 &  1040.3 & -807.5  & 336.8 \\
        0 & -807.5  & 3720.3  & -1551.9 \\
        0 & 336.8   & -1551.9 &  2914.9
        \end{pmatrix}.
    \ee
    \item The DES mission, that publishes the combination $D_M/r_s$ to be 
    \begin{center}
    \begin{tabular}{ |c|c|c|c|c|c|c| } 
    \hline
    Mission & $z$ & $D_M/r_s$ & $\sigma_{D_M/r_s}$ & Ref. & Date\\
    \hline
    DES & $0.835$ & $18.92$ & $0.51$ & \cite{DES:2021esc} & July 2021\\
    \hline
    \end{tabular}
    \end{center}
    which gives a $\chi^2$ formula of the form 
    \be 
    \chi^2_{\rm{DES}}=\sum_{i} \left(\frac{D_M(z,i)/r_s-D_M(z_i)/r_s}{\sigma_{D_M(z,i)/r_s}} \right)^2.
    \ee
    \item The Lya mission which constrains the combinations $f_{BAO}=\left(D_H/r_s,D_M/r_s \right)$ as 
    \begin{center}
    \begin{tabular}{ |c|c|c|c|c|c|c| } 
    \hline
    Mission & $z$ & $f_{BAO}$ & $\sigma_{f_{BAO}}$ & Ref. & Date\\
    \hline
    Lya & $2.33$ & $(8.99,37.5)$ & $(0.19,1.1)$ & \cite{duMasdesBourboux:2020pck} & July 2020\\
    \hline
    \end{tabular}
    \end{center}
    Notice that the $1\sigma$ errors provided in the above table correspond to the  statistical errors. So taking into account the two auto-correlations along with the two cross-correlations as illustrated in Eq. (43) of Ref. \cite{duMasdesBourboux:2020pck} the relevant $\chi^2$ function is constructed as \begin{equation}
    \chi^2_{\rm{Lya}}=V^i \, C_{ij}^{-1} \, V^j.
    \end{equation}
    In this case, the vector $V^i$ reads as $V^i=(f_{BAO}(z,i)-f_{BAO}(z_i))\equiv (D_H(z,i)/r_s-D_H(z_i)/r_s,D_M(z,i)/r_s-D_M(z_i)/r_s)$.
    \item Finally we include the results of the eBOSS mission which contains measurements in various redshift ranges as summarized in detail in Table 3 of Ref. \cite{eBOSS:2020yzd}. Using these values, a $\chi^2$ formula for each datapoint can be constructed of the form 
    \be 
    \chi^2_{\rm{eBOSS}}=V^i \, C_{ij}^{-1} \, V^j,
    \ee
    where the vector $V^i$ corresponds to the difference between the observational values with the respected theoretical expressions and $C_{ij}$ is the corresponding covariance matrix.
    \end{itemize}
Note that some of the individual $\chi^2$ terms might be correlated with each other, e.g. the Lya terms and the ones from eBOSS. Unfortunately however, it is impossible to calculate the possible correlations between the terms as we do not have the necessary covariance matrices, so we will assume they are uncorrelated. Therefore, to obtain the total $\chi^2$, we add all the individual terms as follows \cite{Arjona:2020kco}
\be 
\chi^2_\mathrm{BAO}(r_s\,h, \mathrm{GA})=\chi^2_{\rm{6dFS,WigZ}}+\chi^2_{\rm{DES}}+\chi^2_{\rm{Lya}}+\chi^2_{\rm{eBOSS}} \label{eq:chi2BAO}.
\ee
Furthermore, as mentioned earlier the BAO points depend strongly on the sound horizon at the drag epoch $r_s(z_d)$ times the dimensionless Hubble parameter $h=H_0/(100\,\mathrm{km}\,{s}^{-1}\,\mathrm{Mpc}^{-1})$, i.e. the combination $r_s\,h$. Schematically, this means that the $\chi^2$ is a function of the product of the two variables and the functions produced by the GA, i.e. $\chi^2=\chi^2(r_s\,h, \mathrm{GA})$. However, since we utilize the model-independent GA method the evaluation of this quantity is a rather difficult task.

In order to avoid the dependence on the early time physics of recombination, we then minimize numerically the $\chi^2$ over the quantity $r_s\,h$ and we construct a new $\chi^2$ which is independent of $r_s\,h$, namely: 
\be
\chi^2_\mathrm{BAO}(\mathrm{GA})=\mathrm{min}_{r_s\,h}\Big[\chi^2_\mathrm{BAO}(r_s\,h, \mathrm{GA})\Big].
\ee
We find that this is quite fast and does not affect the efficiency of the GA code. 

Note however, that the minimization procedure used here is effectively using the frequentist profile likelihood, but an arguably more appropriate (albeit more computationally costly approach), would be to  marginalize over this parameter. As the latter would slow down the code significantly, we prefer to use the minimization approach.

\subsection{Cosmic Chronometers}
The CC data correspond to another quite useful probe, since they can directly constrain the Hubble rate $H(z)$ at different redshifts, optimizing the differential age method. In particular the CC measurements are based on the raw definition of the Hubble parameter as 
\be 
H(z)=-\frac{1}{1+z} \, \frac{\mathrm{d}z}{\mathrm{d}t} \label{eq:HzCC},
\ee
avoiding as a result any complex integrations that may appear in other probes such the SnIa or BAO probes. However, it is important to note that the calculation of the ratio $dz/dt$ that appears in the definition \eqref{eq:HzCC} is a rather complex procedure. 

In the present work we use the compilation of 36 $H(z)$ data discussed in Ref. \cite{Arjona:2018jhh} and include 3 additions as it is illustrated in the Table~\ref{tab:data-cc} of Appendix~\ref{sec:dataappdx}. Also, our dataset is similar to that of Ref.~\cite{Moresco:2022phi}, but with the addition of a few more points. Also, here we do not include the full covariance matrix, for which the main effect from that compilation is to just increase the errors in the points and further complicate our analysis.

In any case, using the data of Table~\ref{tab:data-cc}, the standard maximum likelihood method can also be applied here, by constructing a $\chi^2$ function. So, following the method proposed in Ref. \cite{Arjona:2020kco}, we minimize over the $H_0$ parameter (we consider it as a nuisance parameter as we did with $M_B$ in the SnIa case) and find the corresponding $\chi^2$ to be
\be
\chi^2_\mathrm{CC}=A-\frac{B^2}{\Gamma}, \label{eq:chi2CC}
\ee
where the parameters $A,B$ and $\Gamma$ are given by the following equations
\begin{align}
    A &= \sum_i^{N_H} \left(\frac{H_i}{\sigma_{H_i}}\right)^2, \\
    B &= \sum_i^{N_H}\frac{H_i \, E_\textrm{th}(z)}{\sigma_{H_i}^2}, \\
    \Gamma &= \sum_i^{N_H}\left(\frac{E_\textrm{th}(z)}{\sigma_{H_i}}\right)^2,
\end{align}
where $E_{\textrm{th}}(z)$ is defined as $E_{\textrm{th}}(z) \equiv H(z)/H_0$ and $N_H$ describes the total number of CC data. By taking the derivative  of Eq. \eqref{eq:chi2CC} with respect to the nuisance parameter $H_0$ and setting it equal to zero, it is straightforward to show that the minimum is at
\be 
H_0 =\frac{B}{\Gamma}.
\ee
Note that it is in fact easy to show that for any $\chi^2$ that depends quadratically on a parameter, as is the case for the CC data and $H_0$, marginalization of this parameter with a flat prior and direct minimization over it gives the same result, up to an irrelevant constant. Thus, the approach shown here is equivalent to marginalizing over $H_0$.

\subsection{Growth Rate Data}
An additional data compilation that we use in the present work corresponds to the growth rate data compilation presented in Table \ref{tab:data-fs8} of Appendix \ref{sec:dataappdx} (a compilation similar to the one used in Ref.~\cite{Huang:2021tvo}). These data are usually known in the literature as Redshift Space Distortion (RSD) data, due to a particular phenomenon that occurs at both large and small scales during the observation. 

In a nutshell, due to the peculiar velocities of galaxies an overdense region seems squashed in redshift space at large scales, while at small scales an overdense region is elongated along the line of sight affecting as a result the two-point correlation function leading to an anisotropic power spectrum. 

However, on large scales, a part of the observed anisotropy of the power spectrum can also be due to the use of an incorrect fiducial cosmology $H(z)$ that needs to be taken into account when analyzing the growth rate data and corresponds to the so-called Alcock-Paczynski (AP) effect. In the present analysis, we adopt the rough estimate of Ref.~\cite{Kazantzidis:2018rnb}, however many alternative forms have been discussed in the literature.

In order to incorporate the RSD data into our analysis, we move along the lines of Ref.~\cite{Arjona:2020kco} and perform a marginalization process over the constant parameters that appear in $f \sigma_8$ and just rescale the values of the theoretical prediction. If we write $f \sigma_8=\frac{\sigma_8}{\delta_m(1)}\, a\,\delta_m'(a) = f \sigma_{8,0}  \, fs(a) $ where $fs(a)=a\,\delta_m'(a)$, then we can marginalize over the (a priori unknown) scaling constants $f \sigma_{8,0}\equiv \frac{\sigma_8}{\delta_m(1)}$. At this point we also include a correction due to the Alcock-Paczynski (AP) effect, as discussed in Ref.~\cite{Nesseris:2017vor, Kazantzidis:2018rnb}. In particular, the correction can be approximated as \cite{Macaulay:2013swa}
\be 
f\sigma_8(a)\simeq \frac{H(a)D_A(a)}{\tilde{H}(a)\tilde{D_A}(a)}\,\tilde{f\sigma_8}(a),
\ee 
where $H(a)$, $D_A(a)$ are the Hubble parameter and angular diameter distance of the model at hand, $\tilde{H}(a)$, $\tilde{D}_A(a)$ are the Hubble parameter and angular diameter distance of the fiducial cosmology used in the derivation of the data (usually the flat cosmological constant model with some reference $\Omega_\mathrm{m,0}$ value (given for our data in Table \ref{tab:data-fs8}) and $\tilde{f\sigma_8}(a)$ is the reference value of the data, to be corrected. 

While RSD surveys account for the AP effect of each individual point, we still need to correct for it if the cosmology assumed by the survey is quite far from the one where we evaluate $f\sigma_8$. To make matters even more complicated, as seen in Table~\ref{tab:data-fs8}, the fiducial cosmologies assumed by the RSD surveys are quite different from each other. Thus, as it has been noted in the literature, see Ref.~\cite{Macaulay:2013swa} and references there-in, that applying an AP correction again can account (to some extent) for any discrepancies caused by the various fiducial $\Omega_\mathrm{m,0}$ values.

Then we have that $fs(a)=a\,\delta'(a)\, \frac{H(a)D_A(a)}{\tilde{H}(a)\tilde{D_A}(a)}$ and to do the marginalization, we write the $\chi^2$ as 
\bea
\chi^2&=& \left[f \sigma_{8,i}-f \sigma_{8,0} fs(z_i)\right]\,C_{ij}^{-1}\,\left[f \sigma_{8,j}-f \sigma_{8,0} fs(z_j)\right] \nn \\
&=& A-f \sigma_{8,0} B+f \sigma_{8,0}^2 \Gamma,
\eea 
where we have expanded the sum and defined the coefficients $A$, $B$, $\Gamma$ as 
\bea 
A&=& f \sigma_{8,i} \,C_{ij}^{-1}\, f \sigma_{8,j}, \\
B&=& f \sigma_{8,i} \,C_{ij}^{-1}\, fs(z_j)+fs(z_i)\,C_{ij}^{-1}\,f \sigma_{8,j}, \\
\Gamma&=& fs(z_i)\,C_{ij}^{-1}\, fs(z_j).
\eea 
Then the value of $f \sigma_{8,0}$ at the minimum and the corresponding marginalized $\chi^2$ are respectively:
\bea 
f \sigma_{8,0}{}_\mathrm{min}&=&\frac{B}{2\,\Gamma},\\
\tilde{\chi}^2 &=& A-\frac{B^2}{4 \Gamma},
\eea 
which is what we use in our analysis. As mentioned earlier, as the parameter $f \sigma_{8,0}$ appears quadratically in the $\chi^2$, then minimizing over it is in fact equivalent to marginalizing over it.

Finally, even though the RSD and BAO points come in principle from the same or related datasets, we cannot take into account these possible correlations as in general, we have no access to the necessary covariance matrices. Even in the cases of the eBOSS data where we have the covariance between the RSD and BAO point, it is difficult to incorporate the covariance matrix into the analysis as we perform the GA fitting separately. Thus, to keep the analysis simple we will assume we can just add the RSD and BAO $\chi^2$ terms together.

\subsection{$\rm{E_g}$ Data \label{sec:egdata}}
Last but not least, we have used an updated compilation of seven uncorrelated $E_g$ datapoints that are presented in the Table \ref{tab:data-Eg} of Appendix \ref{sec:dataappdx}. These consist of five weak gravitational lensing datapoints from the Kilo-Degree Survey (KiDS-1000), juxtaposed with overlapping data from the BOSS and 2dFLenS galaxy spectroscopic redshift surveys \cite{Blake:2020mzy} and two datapoints from the VIMOS Public Extragalactic Redshift Survey (VIPERS) \cite{delaTorre:2016rxm}. 

The $E_g$ statistic as expressed via Eq.~\eqref{eq:egstat} is galaxy bias-independent at linear order, since it was created by definition as a probe of the ratio of the Newtonian potentials ($\Phi$, $\Psi$) of the perturbed FLRW metric given by Eq.~\eqref{eq:FLRW}. 
In order to fit the $E_g$ data we can construct and minimize a $\chi^2$ formula of the form 
\be
\chi^2_{E_g} = \sum_{i} \left(\frac{2E_g^i - P_2(a^i)}{2\sigma_{E_g, i}}\right)^2.
\ee
Furthermore, in our analysis we use the GA pipeline to directly fit the function $P_2(a)$ given analytically from Eq.~\eqref{eq:P2}. 
 
In general, the $E_g$ data have been known to be plagued by scale and bias-dependent lensing contributions, which in effect increase the systematic uncertainties \cite{MoradinezhadDizgah:2016pqy}. While this can be ameliorated to a certain extend by adding the correlations of shear and galaxy clustering, it is unclear whether the current data have these corrections \cite{Ghosh:2018ijm}.

One possibility to at least take into account this extra uncertainty is to introduce an intrinsic systematic error $\sigma_\mathrm{stat}$ that has to be determined from the data such that the $\chi^2$ per degree of freedom (dof) is order unity, i.e. $\chi^2/\mathrm{d.o.f.}\sim1$, as was done in the past for some SnIa compilations (see Ref.~\cite{SNLS:2005qlf}), but this is a rather questionable statistical practice as it assumes a priori the validity of the model and makes model comparison impossible.

Thus, in what follows we will interpret any results stemming from the currently available $E_g$ data with caution, even if they are very promising and may already weakly hint towards some new physics, as it is uncertain if the possible systematics can explain the observed deviations from GR \cite{Arjona:2020kco}.

\subsection{The mock data \label{sec:mocks}}
Here we briefly discuss the mock data we used in our analysis. In particular, we consider two different fiducial cosmologies, one based on the \lcdm model and another one based on a model with an evolving Newton's constant $\mu(k,a)$ and lensing parameter  $\Sigma(k,a)$, as defined via the Poisson equations Eqs.~\eqref{eq:pois1}-\eqref{eq:pois2}. We call the latter model the $\mu$CDM. Specifically, we use the parametrizations:
\bea 
\mu(k,a) &=& 1 + g_a (1 - a)^{m_1} - g_a (1 - a)^{2 m_1},\label{eq:mock_mu}\\
\Sigma(k,a) &=& 2 + \sigma_a (1 - a)^{m_2} - \sigma_a (1 - a)^{2 m_2},\label{eq:mock_S}
\eea
where $g_a$, $\sigma_a$ are some parameters ($g_a=\sigma_a=0$ in the \lcdm model), and we set $m_1=m_2=2$, inspired by Ref.~\cite{Nesseris:2017vor} such that the models pass the solar system tests. The parameters for the $\mu$CDM are chosen to be $g_a = -0.627$ and $\sigma_a = -3.562$, which are the best-fit values found in Ref.~\cite{Skara:2019usd} (see the ``Datasets $f\sigma_8+E_g$ corr." combination in their Table IV). 

For the rest of the cosmological parameters (which are common to both fiducial cosmologies) we assume $\Omega_\mathrm{m,0}=0.3$, $h=H_0/(100\, \mathrm{km}\,\mathrm{s}^{-1}\,\mathrm{Mpc}^{-1})=0.7$, $\sigma_{8,0}=0.8$ and for the DE parameters $(w_0,w_a)=(-1,0)$, i.e. we assume a \lcdm background as we are interested in the effects of the cosmological perturbations.

\begin{figure*}[!t]
\centering
\includegraphics[width = 0.49\textwidth]{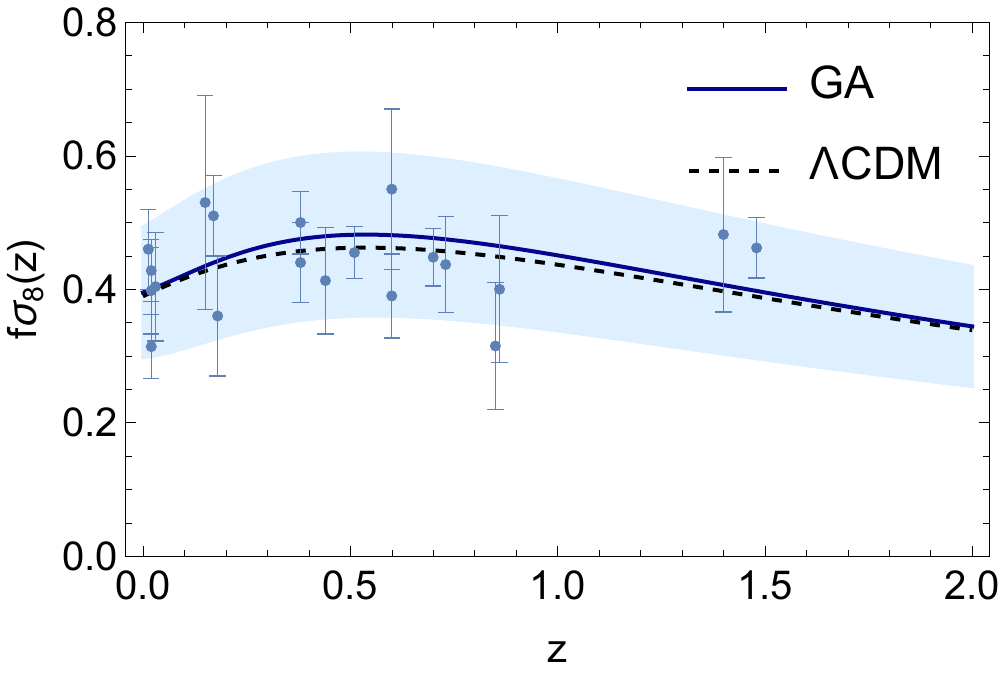}
\includegraphics[width = 0.49\textwidth]{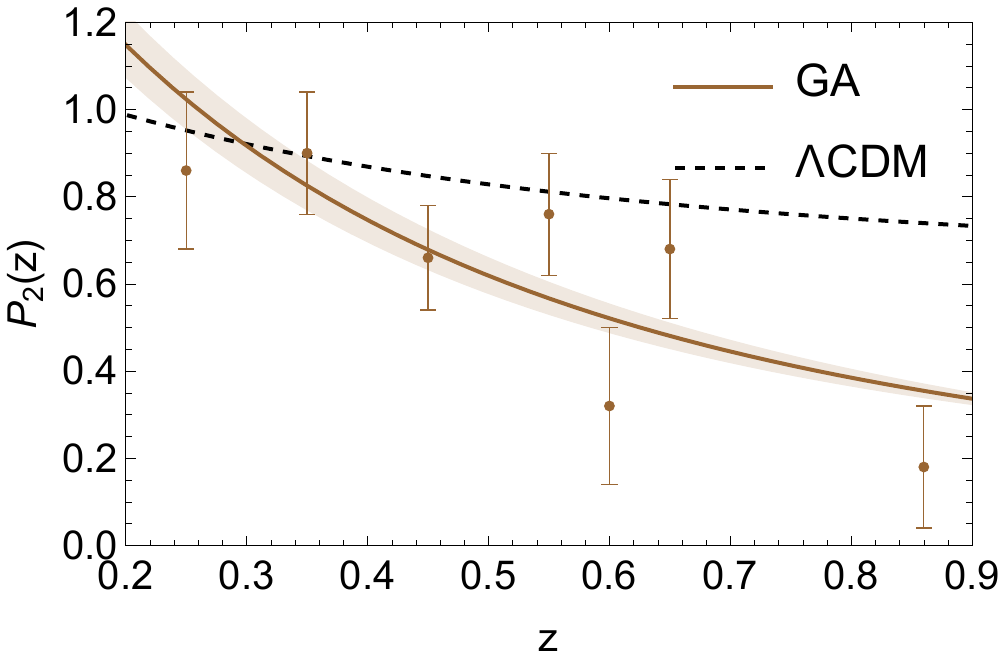}
\caption{The best-fit $f\sigma_8(z)$ (left) and $P_2(z)$ (right) functions (blue and brown lines), with their respective datapoints, as predicted by the GA pipelines using growth and $E_g$ data accordingly, along with their $1\sigma$ error bands (light blue and light brown areas). We see that while on one hand, the predicted $f\sigma_8(z)$ function is consistent with the one produced assuming \lcdm (dashed black line) the same cannot be said for the predicted $P_2(z)$ function, which is affected by the low quality of the data.}
\label{fig:fs8P2plt}
\end{figure*}

\begin{figure*}[!t]
\centering
\includegraphics[width = 0.47\textwidth]{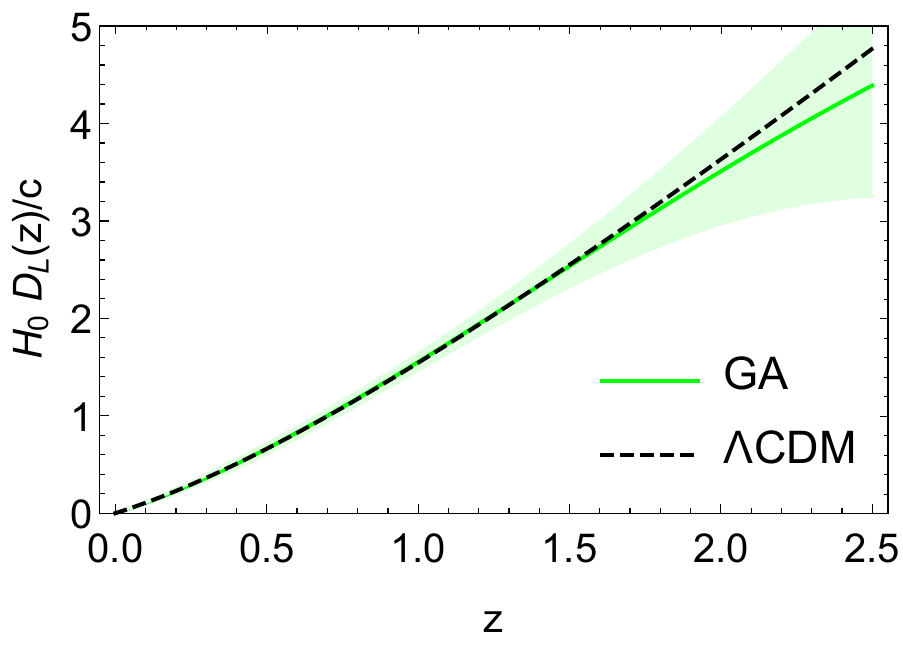}
\includegraphics[width = 0.49\textwidth]{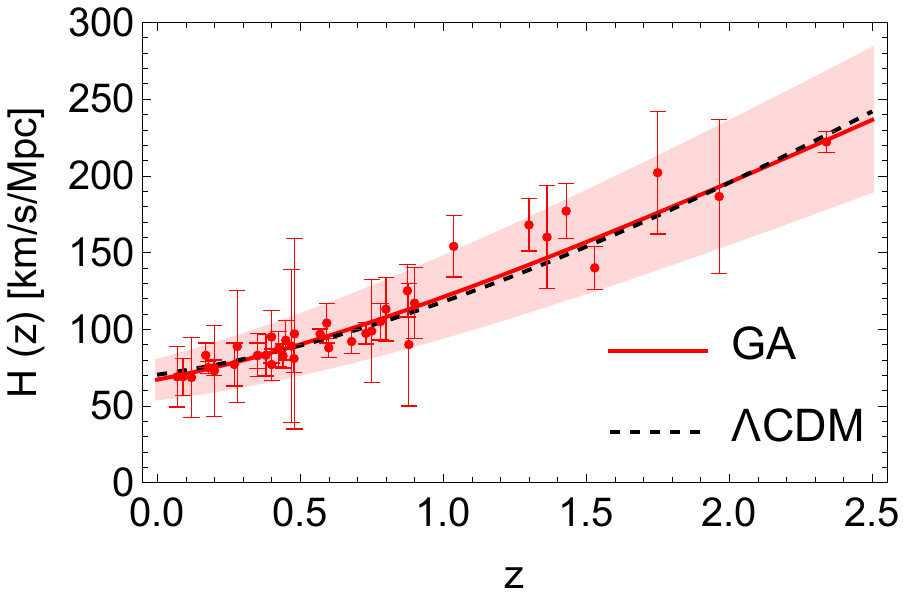}
\caption{The best-fit for the luminosity distance $D_L(z)$ (left) and Hubble rate $H(z)$ (right) functions (green and red lines) as predicted by the GA using CC and SnIa data accordingly, with the CC datapoints, along with their $1\sigma$ error bands (light green and light red areas). The dashed black line corresponds to the \lcdm best-fit.}
\label{fig:dLHzplt}
\end{figure*}

Having fixed the cosmological parameters for the two models, we then make the corresponding two sets of data for the Hubble parameter $H(z)$, the growth rate $f\sigma_8(z)$ and the $E_g(z)$ parameter for each model in twenty redshift bins in the range $z\in[0,2.0]$ with a fixed bin size of $dz=0.1$. We also add Gaussian noise and errors to the $1\%$ level, which is within the expectations of forthcoming next-generation surveys, such as Euclid  \cite{EUCLID:2011zbd}.

In fact, Euclid will obtain precise measurements of these quantities, albeit its redshift range is much more limited, e.g. for the spectroscopic survey the range is $z\in [0.9,1.8]$ and needs to be complemented by other surveys (e.g. DESI) in lower redshifts \cite{Euclid:2021frk,Euclid:2021cfn,EUCLID:2020syl}. Thus, we consider our methodology as a proof of concept approach and the forecasts as an optimistic case to gauge how well the method works overall.

With these two mock data sets at hand, we then rerun the GA pipelines as in the previous section and we again reconstruct the parameters $\mu(k,a$) and $\eta(k,a)$, aiming to forecast at which level our ML approach is able to detect deviations from GR. One  key difference from the analysis of the real data in  the previous section is though that in this case we do not require any SnIa data or any corrections for the AP effect in the growth data, as all the points will come from the same source, unlike the compilation of the currently available points used here, thus allowing us to perform a much cleaner analysis.

We discuss the results of our analysis in the next section in detail. 

\section{Results \label{sec:results}}
Using the methodology described earlier, in order to directly reconstruct $\mu(z)$ and $\eta(z)$ we aim to produce analytic fits for the functions $H(z)$, $D_{L}(z)$, $D_{A}(z)$, $f\sigma_{8}(z)$ and $E_g(z)$, via updated compilations of the CC, SnIa, BAO, RSD and $E_g$ data  for the real data. On the other hand,  fit for the functions $H(z)$, $f\sigma_{8}(z)$ and $E_g(z)$ in the case of the mock data.

Before running the GA analysis we choose a specific grammar and we also impose a few physically motivated hypotheses in order to ensure that the obtained functions are well-behaved, e.g. smooth, continuous and singularity-free as expected by physical quantities. In particular, we impose the following priors:
\begin{itemize}
    \item The luminosity distance $D_L(z)$ in the low redshift regime should be approximated by the Hubble law as $D_L(z \rightarrow 0) \simeq c \, z/ H_0$. The same behaviour is also expected for the angular diameter distance $D_A(z)$ in the low redshift regime, since the two quantities in the context of FLRW metric are connected via the standard distance duality relation 
    \be 
    D_L(z)=(1+z)^2 \, D_A(z).
    \ee
    \item The present-day value of the Hubble parameter should be equal to the Hubble constant $H_0$, which is evaluated via the CC data.
    \item In the high redshift regime, i.e. deep in the matter domination era, the growth factor $\delta_m(a)$ should evolve as $\delta_m(a) \simeq a$, as expected from LSS observations. 
\end{itemize}
The particular prior for the distances is implemented by demanding that the luminosity distance behaves as $D_L(z)=c\,H_0^{-1}\,z\,\big[1+z\cdot\mathrm{GA}(z)\big]$, where $\mathrm{GA}(z)$ is the function predicted by the GA. This ensures that the prior is actually enforced at all times for well-behaved functions (as are the ones created by our implementation of the GA). Also, in order to maximize the freedom in the reconstructions in this work we follow Ref.~\cite{Arjona:2020kco} and perform a separate analysis by using two separate GA functions for $D_L(z)$ and $H(z)$. 

However, for completeness we also compared the expressions for the luminosity distance: one determined directly by the GA and the other by integrating the Hubble parameter. Doing so we find that in the redshift range $z\in[0,2]$ the agreement between the two approaches is on average $\sim1.2\%$ and always below $2\%$. Thus, there is reasonable agreement between the two function, even if we do not enforce Eq.~\eqref{eq:dLt}.

On the other hand, in the high redshift regime, i.e. deep in the matter domination era, while the growth factor $\delta_m(a)$ should evolve as $\delta_m(a)\simeq a$, as expected from LSS observations, these high redshifts ($z\in[10^2,10^4]$) are not yet probed by LSS observations and are only indirectly constrained by the CMB, so any deviations from the expected behavior are only speculative at the moment. Still, in order to probe for any deviations, we extend the GA grammar used, by considering also polynomials of fractional powers of the scale factor $a$.

Finally, the initial functions are then subjected to the crossover and mutation operations and after this process is repeated several times (until we reach a maximum number of generations to ensure convergence), the final expressions for the best-fit functions are obtained.

\subsection{Currently available data}
\begin{figure}[!t]
\centering
\includegraphics[width = 0.49\textwidth]{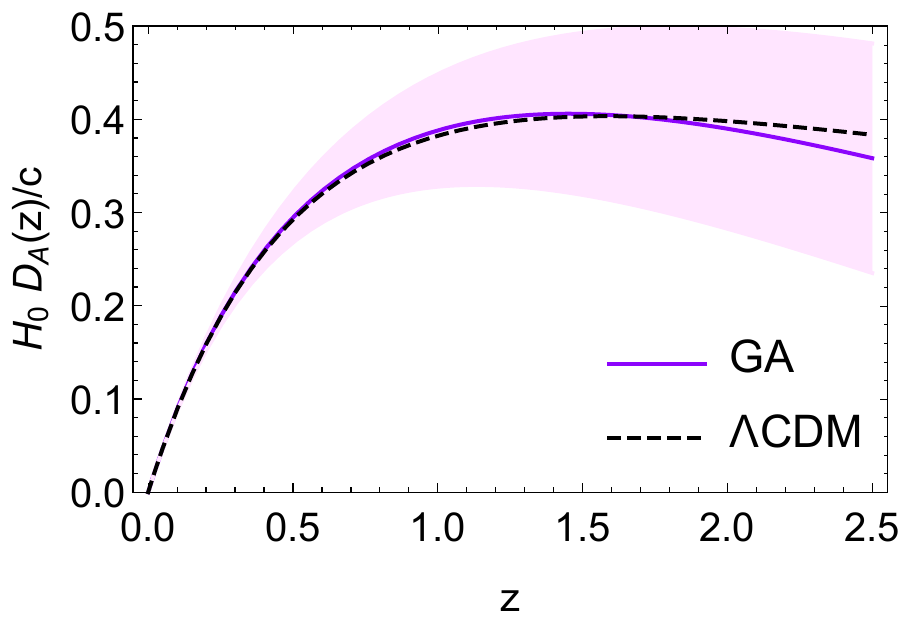}
\caption{The best-fit angular diameter distance $D_A(z)$ function (magenta line) as predicted by the GA using a compilation of BAO data, along with its $1\sigma$ error band (light magenta area). The dashed black line corresponds to the \lcdm best-fit. }\label{fig:daplt}
\end{figure}

Running the GA pipelines for the currently available data, the following best-fit function forms are obtained:
\begin{widetext}
\begin{align}
    \left(\frac{H(z)}{H_0}\right)^2 &= 1+z \left[0.262 \, (0.153z+1)^3-0.618\,z -1.281\right]^2,\\
    D_L(z) &=\frac{c}{H_0} z \left[1+z \, (0.872\, -0.129 \, z)^2\right], \\
    f\sigma_8(a) &= f_0 \, \left \lbrace a-a^4 \left[1.680 \left(0.073^{0.073 a} a^{0.073 a}\right)^{1.720}+0.003 a^{5.680}-0.412 a^{1.533}\right]^2 \right\rbrace, \\
    P_2(a) &= \left(-0.160 \cdot 0.180^{0.361 a} a^{0.361 a}+4.758 a^9-2.869 a^3\right)^2,
\end{align}
\end{widetext}
where $f_0=1.109\pm0.272$ and $H_0=(67.17\pm 12.22)\, \mathrm{km}\,\mathrm{s}^{-1}\,\mathrm{Mpc}^{-1}$, which is compatible with the Planck 2018 measurements. The redshift evolution of the growth rate $f\sigma_8(z)$ and of the quantity $P_2(z)$ as reconstructed by the GA using the RSD and $E_g$ data, is illustrated in Fig.~\ref{fig:fs8P2plt} (blue and brown lines), while the evolution of the Hubble rate $H(z)$, of the luminosity distance $D_L(z)$ and of the angular diameter distance $D_A(z)$ as reconstructed by the GA using the CC, the BAO and the SnIa data, is illustrated in Figs.~\ref{fig:dLHzplt} and \ref{fig:daplt} accordingly (red, green and magenta lines). In each plot we also include the best-fit of the \lcdm scenario (dashed black lines) as well as the $1\sigma$ errors regions (shaded areas) utilizing the path integral approach.

Clearly, the \lcdm case is consistent with the results of the GA, \ie the black dashed line is allocated well within $\sim1\sigma$ in almost all of the cases except the results produced from the $E_g$ data, where the $P_2(z)$ function as derived by the GA pipeline presents a large deviation from the one predicted for the \lcdm case. However, it is important to note that even the \lcdm scenario does not fit well the $E_g$ data compilation, the fact that the $\chi^2$ per degree of freedom is $\simeq 1.8$ is an indication that the $E_g$ results should be interpreted with care. Interestingly, a similar deviation from the \lcdm case was also observed for $P_2(z)$ in Ref. \cite{Arjona:2020kco}.

\begin{figure*}
\centering
\includegraphics[width = 0.47\textwidth]{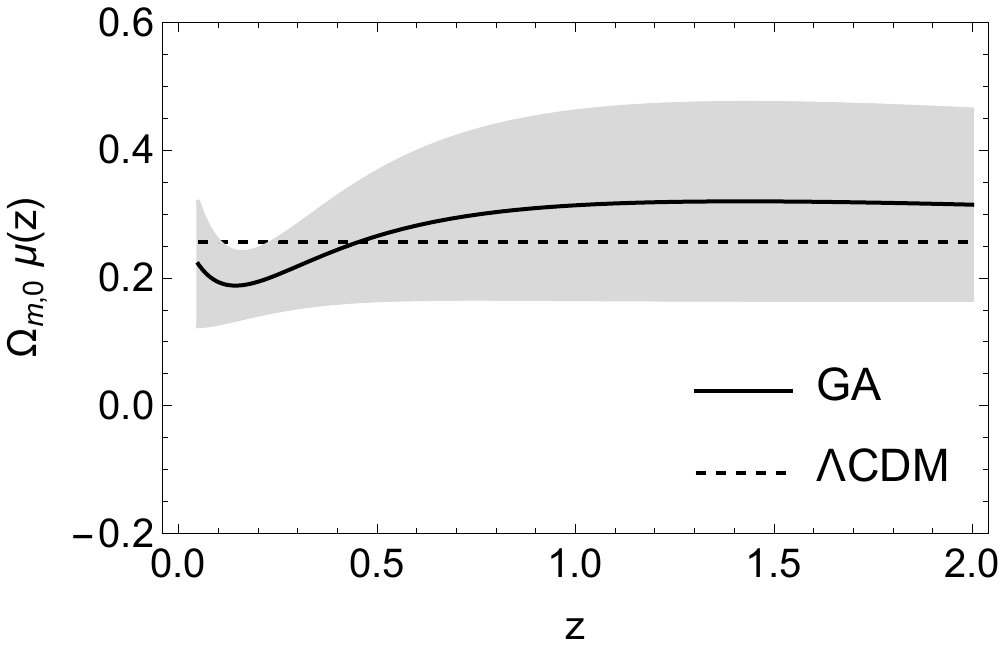}
\includegraphics[width = 0.48\textwidth]{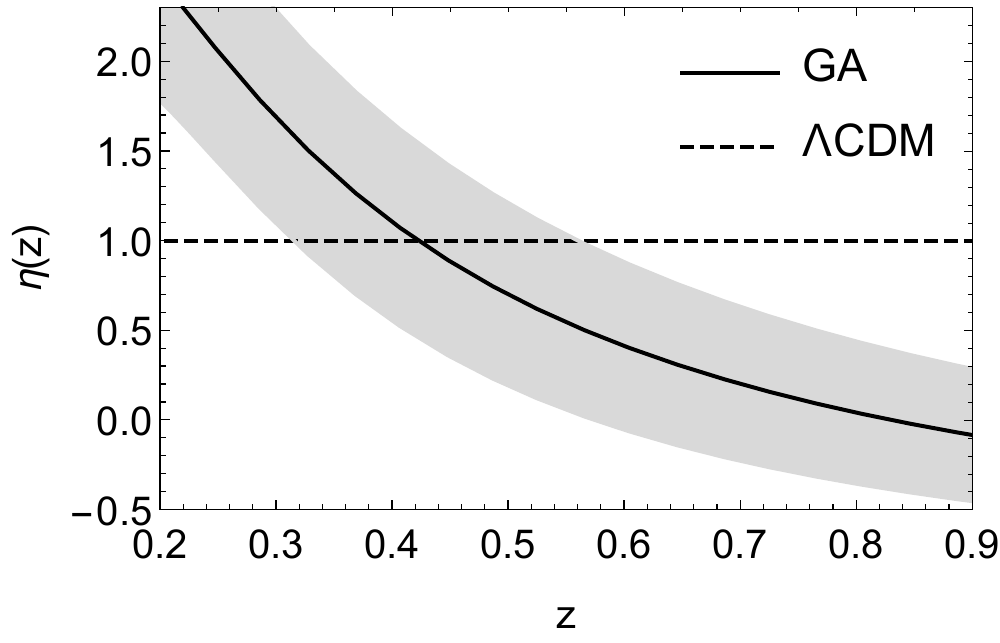}
\caption{The reconstructions of $\Omega_\mathrm{m,0}\,\mu(z)$ (left) and $\eta(z)$ (right) given by Eq.~\eqref{eq:mufrac} and \eqref{eq:eta} respectively, where the relevant functions were produced by the GA pipelines. The solid black lines correspond to the GA reconstruction, while the $1\sigma$ error band is the light gray area. Note that we only show the reconstruction of $\eta$ in the redshift range $z\in [0.2,0.9]$ where we have the $E_g$ data.\label{fig:muplt}}
\end{figure*}

Using the aforementioned GA reconstructions we finally calculate the evolution of the parameter $\Omega_\mathrm{m,0}\,\mu(z)$ using the theoretical expressions presented in Sec.~\ref{sec:theory}. The final result is shown in the left panel of Fig.~\ref{fig:muplt} where we show the GA reconstruction in a solid black line along with the $1\sigma$ errors as the shaded region. Overall we find that while there is good agreement with the expectation of a constant value, but we find that the errors are quite large due to the lower quality of the currently available data compared to the ones from forthcoming surveys in the near future, thus not allowing us to draw any strong conclusions on any deviations of $\Omega_\mathrm{m,0}\,\mu(z)$ from a constant value.

We also perform the GA reconstruction of the anisotropic stress $\eta$, which is shown in the right panel of Fig.~\ref{fig:muplt} in the redshift range $z\in [0.2,0.9]$ where we have the $E_g$ data, but we find that it deviates strongly from unity. The reason for this is that the $E_g$ data are most likely plagued with systematics (as discussed in detail in Sec.~\ref{sec:egdata}) which most likely drive the deviation seen. Of course, the possibility of new physics also cannot be excluded as the systematics in the data are unlikely to account for the whole deviation from unity observed in $\eta$. A similar result was found for $\eta$ in Ref.~\cite{Arjona:2020kco} using an earlier compilation of the data.

\subsection{Mock Data}

\begin{figure*}[!t]
\centering
\includegraphics[width = 0.495\textwidth]{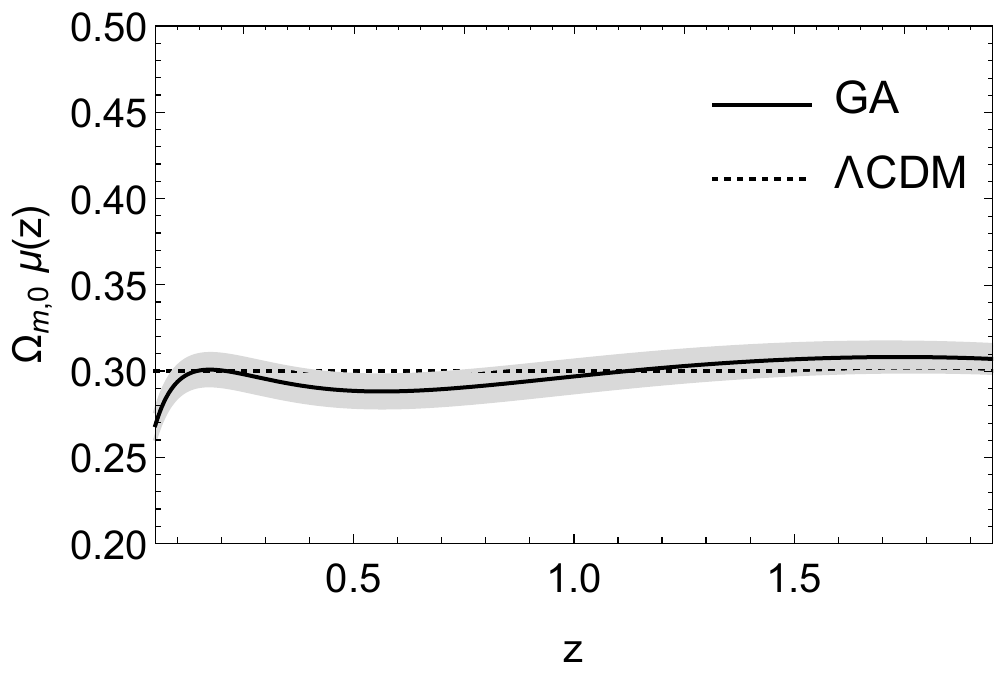}
\includegraphics[width = 0.495\textwidth]{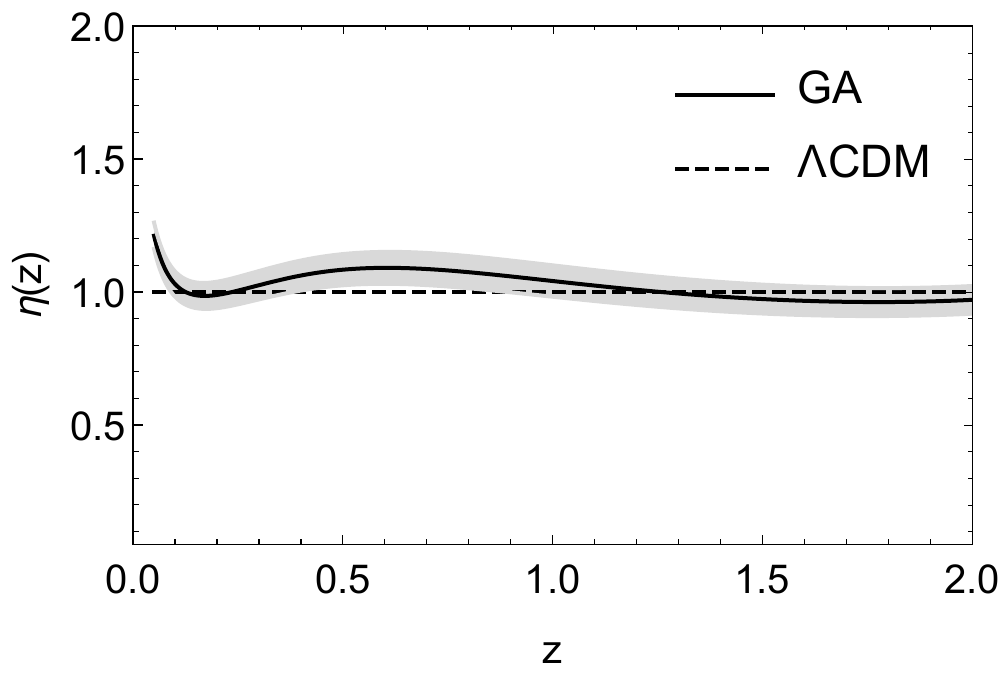}
\caption{The GA reconstruction of $\Omega_\mathrm{m,0}\,\mu(z)$ (left) and $\eta(z)$ (right) for the \lcdm mock. The gray-shaded regions correspond to the $1\sigma$ confidence level, while the dashed red line corresponds to the fiducial model $\mu(z)=\eta(z)=1$, i.e. GR and the \lcdm model. In both cases the GA is able to constraint the correct underlying fiducial model to within a few percent.\label{fig:mu_eta_lcdm}}
\end{figure*}

\begin{figure*}[!t]
\centering
\includegraphics[width = 0.495\textwidth]{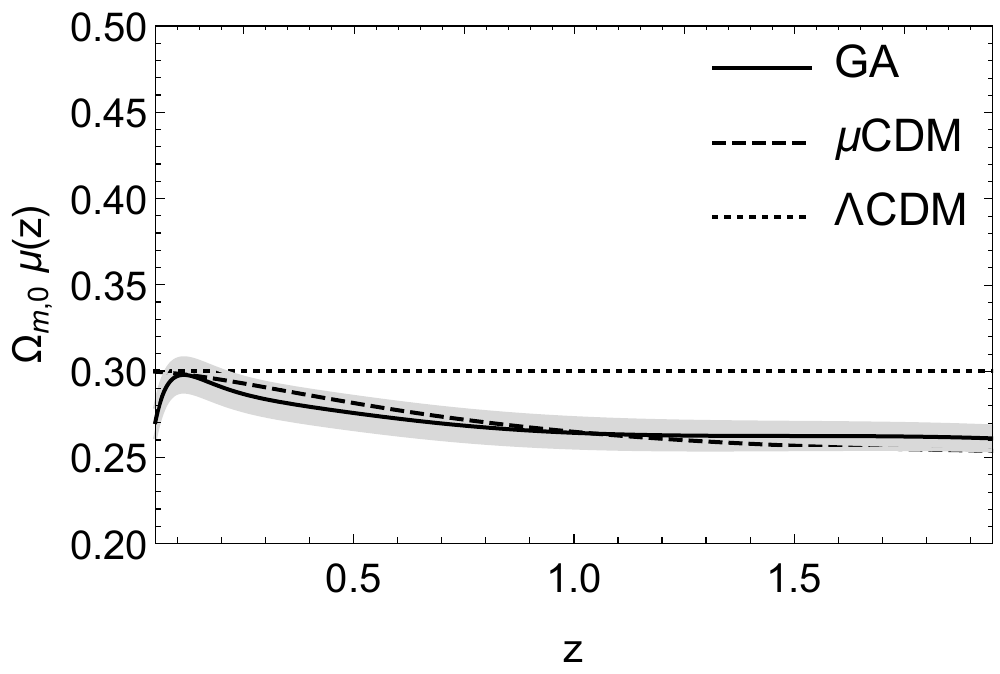}
\includegraphics[width = 0.495\textwidth]{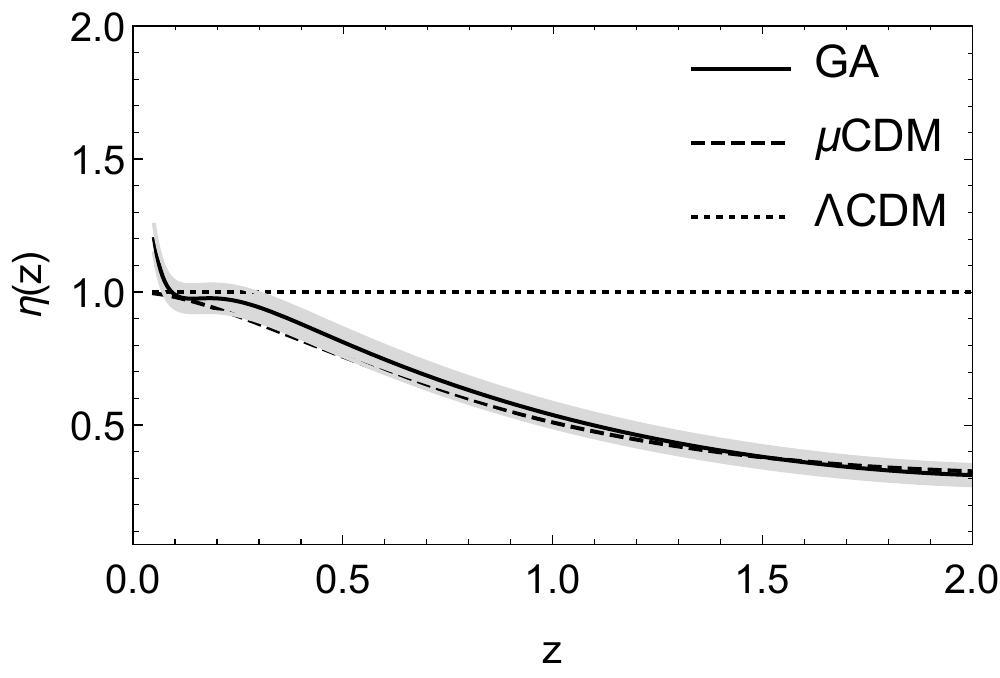}
\caption{The GA reconstruction of $\Omega_\mathrm{m,0}\,\mu(z)$ (left) and $\eta(z)$ (right) for the $\mu$CDM mock. The gray-shaded regions correspond to the $1\sigma$ confidence level, while the dashed red line corresponds to the fiducial model described by Eqs.~\eqref{eq:mock_mu}-\eqref{eq:mock_S} and the parameters $g_a = -0.627$ and $\sigma_a = -3.562$ for $m_1=m_2=2$.
\label{fig:mu_eta_mucdm}}
\end{figure*}

Here we repeat the GA reconstruction analysis, however this time using the mock data  described in Sec.~\ref{sec:mocks}. In particular we run the whole GA pipeline on the \lcdm and $\mu$CDM mocks and from there we reconstruct the quantity $\Omega_\mathrm{m,0}\,\mu(z)$ which is related to the effective Newton's constant and the anisotropic stress parameter $\eta(z)$ using Eqs.~\eqref{eq:om_mu} and \eqref{eq:eta_obs} respectively.

In Fig.~\ref{fig:mu_eta_lcdm} we show the results of these reconstructions for the \lcdm mock. In particular, we show $\Omega_\mathrm{m,0}\,\mu(z)$ in the left panel and $\eta(z)$ on the right panel of Fig.~\ref{fig:mu_eta_lcdm} respectively and as can be seen, in both cases the GA is able to recover the correct underlying cosmology within the errors, given by the gray-shaded regions at the $1\sigma$ confidence level.

On the other hand, in Fig.~\ref{fig:mu_eta_mucdm} we show the GA reconstructions for the $\mu$CDM mock, again $\Omega_\mathrm{m,0}\,\mu(z)$ in the left panel and $\eta(z)$ on the right panel respectively. Here we find that even though the fiducial cosmology is quite extreme, the GA is able to reconstruct both functions to within a few percent and provide a detection, quantified as a deviation from a constant value, at several $\sigma$s.

\section{Conclusions \label{sec:conclusions}}
In this work used a particular ML approach, called the GA, in order to perform non-parametric reconstructions of two key quantities that parameterize deviations from GR, namely the effective Newton's constant $G_\mathrm{eff}$ and the anisotropic parameter $\eta$ \cite{Sobral-Blanco:2021cks, Tutusaus:2022cab}.

To perform the reconstructions we used both the currently available data, coming from the BAO, the CC, the growth of matter perturbations and the so-called $E_g$ statistics compilations, but also synthetic data assuming a next generation survey in order to forecast, in an optimist scenario, how well we will be able to constrain deviations from the null hypothesis, either the \lcdm model or a model with an evolving Newton's constant, dubbed $\mu$CDM.

In the case of the currently available data we find, as expected, that the GA reconstructions are affected by the lower quality of the currently available data. While $\Omega_\mathrm{m,0}\,\mu(z)$ seems to be in good agreement with having a constant value at all redshifts within the erros, on the other hand the anisotropic stress $\eta$ is more difficult to interpret as the fit is plagued by the systematics of the $E_g$ data, in agreement with previous works. While it might have been interesting to repeat the same analysis without the $E_g$ data, it is in fact not possible to do that as $\Omega_\mathrm{m,0}\,\mu(z)$ does not contain the $E_g$ function but $\eta(z)$ requires it, thus we cannot do a separate analysis.

The situation is more clear when using the fiducial data based on the two different mocks. Here we find that in both cases the GA reconstructions are in good agreement with the fiducial models, either the \lcdm for which $\mu=\eta=1$ or the $\mu$CDM model which had a more complicated evolution. 

In particular, we see in Figs.~\ref{fig:mu_eta_lcdm}-\ref{fig:mu_eta_mucdm} that the GA reconstructions are well within the $1\sigma$ errors of the true fiducial model and the GA is able to constrain the correct underlying fiducial model to within a few percent. This is possible for both mocks, and in the case of the $\mu$CDM mock, which was based on a rather extreme cosmology, can provide a strong detection of several $\sigma$s.

Overall, we find that this reconstruction method is a very useful tool in performing model-independent reconstructions of the two key quantities $\Omega_\mathrm{m,0}\,\mu(z)$ and $\eta(z)$ that can be used to parameterize most modified gravity models and for which any deviations from unity would be smoking guns for new physics if the systematics are under control.

\textit{Numerical Analysis Files}: The \texttt{Mathematica} codes used by the authors in the analysis of the paper will be made publicly available upon publication at \href{https://github.com/snesseris/GA-Geff-GR}{https://github.com/snesseris/GA-Geff-GR} 

\section*{Acknowledgements}
The authors are grateful to S.~Kuroyanagi and L.~Perivolaropoulos for useful discussions and feedback on the draft. SN acknowledges support from the research project PGC2018-094773-B-C32. 
GA's research is supported by the project “Dioni: Computing Infrastructure for Big-Data Processing and Analysis” (MIS No. 5047222) co-funded by European Union (ERDF) and Greece through Operational Program “Competitiveness, Entrepreneurship and Innovation”, NSRF 2014-2020 and by the Spanish Attraccion de Talento contract no. 2019-T1/TIC-13177 granted by Comunidad de Madrid. 
Both SN and GA are also supported by the Spanish Research Agency (Agencia Estatal de Investigaci\'on) through the Grant IFT Centro de Excelencia Severo Ochoa No CEX2020-001007-S, funded by MCIN/AEI/10.13039/501100011033.

\appendix

\section{Data Compilations \label{sec:dataappdx}}
We present the updated compilations of cosmological data that were used in the analysis.

\begin{widetext}
\begin{longtable}[c]{|c|c|c|c|c|}
\caption{A compilation of 39 Cosmic Chronometer data dating from 2009 to 2022, in units of $\mathrm{km}\,\mathrm{s}^{-1}\,\mathrm{Mpc}^{-1}$.}
\label{tab:data-cc}\\
\hline
   Index & $z$ & $H(z)$ $\pm \rm{\sigma_{\textrm{H}}}$ & Ref. & Date \\
\hline   
1  &  $0.09$    & $69 \pm 12$       & \cite{Stern:2009ep} & July 2009\\
2  &  $0.17$    & $83 \pm 8$        & \cite{Stern:2009ep} & July 2009\\
3  &  $0.27$    & $77 \pm 14$       & \cite{Stern:2009ep} & July 2009\\
4  &  $0.40$    & $95 \pm 17$       & \cite{Stern:2009ep} & July 2009\\
5  &  $0.48$    & $97 \pm 62$       & \cite{Stern:2009ep} & July 2009\\
6  &  $0.88$    & $90 \pm 40$       & \cite{Stern:2009ep} & July 2009\\
7  &  $0.90$    & $117 \pm 23$      & \cite{Stern:2009ep} & July 2009\\
8  &  $1.30$    & $168 \pm 17$      & \cite{Stern:2009ep} & July 2009\\
9  &  $1.43$    & $177 \pm 18$      & \cite{Stern:2009ep} & July 2009\\
10 &  $1.53$    & $140 \pm 14$      & \cite{Stern:2009ep} & July 2009\\
11 &  $1.75$    & $202 \pm 40$      & \cite{Stern:2009ep} & July 2009\\
12 &  $0.44$    & $82.6 \pm 7.8$    & \cite{Blake:2012pj} & June 2012\\
13 &  $0.60$    & $87.9 \pm 6.1$    & \cite{Blake:2012pj} & June 2012\\
14 &  $0.73$    & $97.3 \pm 7.0$    & \cite{Blake:2012pj} & June 2012\\
15 &  $0.179$   & $75 \pm 4$        & \cite{Moresco:2012jh} & February 2013\\
16 &  $0.199$   & $75.0 \pm 5$      & \cite{Moresco:2012jh} & February 2013\\
17 &  $0.352$   & $83.0 \pm 14$     & \cite{Moresco:2012jh} & February 2013\\
18 &  $0.593$   & $104.0 \pm 13$    & \cite{Moresco:2012jh} & February 2013\\
19 &  $0.68$    & $92.0 \pm 8$      & \cite{Moresco:2012jh} & February 2013\\
20 &  $0.781$   & $105.0 \pm 12$    & \cite{Moresco:2012jh} & February 2013\\
21 &  $0.875$   & $125.0 \pm 17$    & \cite{Moresco:2012jh} & February 2013\\
22 &  $1.037$   & $154.0 \pm 20$    & \cite{Moresco:2012jh} & February 2013\\
23 &  $0.35$    & $82.7 \pm 8.4$    & \cite{Chuang:2012qt}  & August 2013\\
24 &  $0.07$    & $69.0 \pm 19.6$   & \cite{Zhang:2012mp} & May 2014\\
25 &  $0.12$    & $68.6 \pm 26.2$   & \cite{Zhang:2012mp} & May 2014\\
26 &  $0.20$    & $72.9 \pm 29.6$   & \cite{Zhang:2012mp} & May 2014\\
27 &  $0.28$    & $88.8 \pm 36.6$   & \cite{Zhang:2012mp} & May 2014\\
28 &  $0.57$    & $96.8 \pm 3.4$    & \cite{BOSS:2013rlg} & June 2014\\
29 &  $2.34$    & $222.0 \pm 7.0$   & \cite{BOSS:2014hwf} & December 2014\\
30 &  $1.363$   & $160.0 \pm 33.6$  & \cite{Moresco:2015cya} & March 2015\\
31 &  $1.965$   & $186.5 \pm 50.4$  & \cite{Moresco:2015cya} & March 2015\\
32 &  $0.3802$  & $83.0 \pm 13.5$   & \cite{Moresco:2016mzx} & May 2016\\
33 & $0.4004$   & $77.0 \pm 10.2 $  & \cite{Moresco:2016mzx} & May 2016\\
34 & $0.4247$   & $87.1 \pm 11.2$   & \cite{Moresco:2016mzx} & May 2016\\
35 & $0.4497$  & $92.8 \pm 12.9$   & \cite{Moresco:2016mzx} & May 2016\\
36 & $0.4783$   & $80.9 \pm 9.0$    & \cite{Moresco:2016mzx} & May 2016\\
37 & $0.47$     & $89 \pm 50$   & \cite{Ratsimbazafy:2017vga}  & February 2017\\
38 & $0.75$     & $98.8 \pm 33.6$   & \cite{Borghi:2021rft}  & October 2021\\
39 & $0.80$     & $113.1 \pm 20.73$  & \cite{Jiao:2022aep} & May 2022  \\
\hline
\end{longtable}
\end{widetext}

\begin{widetext}
\begin{longtable}[c]{|c|c|c|c|c|c|c|}
\caption{An updated compilation of RSD data. The correlations pertaining to the eBOSS data at $z=0.38$ and $z=0.51$ are explained \href{https://svn.sdss.org/public/data/eboss/DR16cosmo/tags/v1_0_0/likelihoods/BAO-plus/}{here}.}
\label{tab:data-fs8}\\
\hline
   Index & Survey & $z$ & $f \sigma_{8} \pm \sigma_{f \sigma_{8}}$ & $\Omega_\mathrm{m,0}$ & Ref. & Date \\
\hline   
1  &  2dFGRS & $0.17$ & $0.510 \pm 0.060$ & $0.3$ & \cite{Song:2008qt} & July 2008 \\
2  &  2MASS & $0.02$ & $0.314 \pm 0.048$ & $0.266$ & \cite{Davis:2010sw,Hudson:2012gt} & January 2011 \\
3  &  SnIa+IRAS & $0.02$ & $0.398 \pm 0.065$ & $0.3$ & \cite{Turnbull:2011ty,Hudson:2012gt} & November 2011 \\
4  &  WiggleZ & $0.44$ & $0.413 \pm 0.080$ & $0.27$ & \cite{Blake:2012pj} & April 2012 \\
5  &  WiggleZ & $0.60$ & $0.390 \pm 0.063$ & $0.27$ & \cite{Blake:2012pj} & April 2012 \\
6  &  WiggleZ & $0.73$ & $0.437 \pm 0.072$ & $0.27$ & \cite{Blake:2012pj} & April 2012 \\
7  &  GAMA & $0.18$ & $0.36 \pm 0.09$ & $0.27$ & \cite{Blake:2013nif} & September 2013 \\
8  &  GAMA & $0.38$ & $0.44 \pm 0.06$ & $0.27$ & \cite{Blake:2013nif} & September 2013 \\
9  & FastSound & $1.4$ & $0.482\pm0.116$ & $0.27$ & \cite{Okumura:2015lvp} & March 2016 \\
10  &  6dFGS+SnIa & $0.02$ & $0.428_{-0.045}^{+0.048}$ & $0.3$ & \cite{Huterer:2016uyq} & November 2016 \\
11 &  VIPERS PDR2 & $0.6$ & $0.55 \pm 0.12$ & $0.3$ & \cite{Pezzotta:2016gbo} & December 2016 \\
12 &  VIPERS PDR2 & $0.86$ & $0.40 \pm 0.11$ & $0.3$ & \cite{Pezzotta:2016gbo} & December 2016 \\
13  & 2MTF+6dFGSv & $0.03$&$0.404^{+0.082}_{-0.081}$ &$0.312$ & \cite{Qin:2019axr} & June 2019 \\
14  & ALFALFA & $0.013$ & $0.46 \pm 0.06$ & $0.315$ & \cite{Avila:2021dqv} & May 2021 \\
15  & eBOSS & $0.15$ & $0.53 \pm 0.16$ & $0.31$ & \cite{eBOSS:2020yzd} & July 2020 \\
16  & eBOSS & $0.38$ & $0.500 \pm 0.047$ & $0.31$ & \cite{eBOSS:2020yzd} & July 2020 \\
17  & eBOSS & $0.51$ & $0.455 \pm 0.039$ & $0.31$ & \cite{eBOSS:2020yzd} & July 2020 \\
18  & eBOSS & $0.70$ & $0.448 \pm 0.043$ & $0.31$ & \cite{eBOSS:2020yzd} & July 2020 \\
19  & eBOSS & $0.85$ & $0.315 \pm 0.095$ & $0.31$ & \cite{eBOSS:2020yzd} & July 2020 \\
20  & eBOSS & $1.48$ & $0.462 \pm 0.045$ & $0.31$ & \cite{eBOSS:2020yzd} & July 2020 \\
\hline
\end{longtable}
\end{widetext}

\begin{widetext}
\begin{longtable}[c]{|c|c|c|c|c|c|}
\caption{The updated and uncorrelated compilation of $\rm{E_g}$ data from 2016 to 2020.}
\label{tab:data-Eg}\\
\hline
   Index & Survey & $z$ & $\rm{E_g} \pm \rm{\sigma_{E_g}}$ & Ref. & Date \\
\hline   
1  &  VIPERS & $0.6$ & $0.16 \pm 0.09$ & \cite{delaTorre:2016rxm} & December 2016 \\
2  &  VIPERS & $0.86$ & $0.09 \pm 0.07$ & \cite{delaTorre:2016rxm} & December 2016 \\
3  &  KiDS-1000+BOSS DR12+2dFLenS & $0.25$ & $0.43 \pm 0.09$ & \cite{Blake:2020mzy} & May 2020 \\
4  &  KiDS-1000+BOSS DR12+2dFLenS & $0.35$ & $0.45 \pm 0.07$ & \cite{Blake:2020mzy} & May 2020 \\
5  &  KiDS-1000+BOSS DR12+2dFLenS & $0.45$ & $0.33 \pm 0.06$ & \cite{Blake:2020mzy} & May 2020 \\
6  &  KiDS-1000+BOSS DR12+2dFLenS & $0.55$ & $0.38 \pm 0.07$ & \cite{Blake:2020mzy} & May 2020 \\
7  &  KiDS-1000+BOSS DR12+2dFLenS & $0.65$ & $0.34 \pm 0.08$ & \cite{Blake:2020mzy} & May 2020 \\
\hline
\end{longtable}
\end{widetext}


\raggedleft
\hypersetup{linkcolor=red}
\bibliography{biblio}

\end{document}